\documentclass[journal]{IEEEtran}
\usepackage{url}
\usepackage[utf8]{inputenc}
\usepackage{xcolor}
\usepackage{amsmath}
\usepackage{amssymb}

\usepackage[acronyms,nonumberlist,nopostdot,nomain,nogroupskip]{glossaries}
\usepackage{tablefootnote}
\usepackage{booktabs}
\usepackage{tabularx}
\usepackage{tikz}
\usepackage{pgfplots}
\pgfplotsset{compat=newest} 
\pgfplotsset{plot coordinates/math parser=false} 
\newlength\fheight
\newlength\fwidth
\usetikzlibrary{plotmarks,patterns,decorations.pathreplacing,backgrounds,calc,arrows,arrows.meta,spy,matrix}
\usepgfplotslibrary{patchplots,groupplots}
\usepackage{tikzscale}
\usepackage{hyperref}
\usepackage{siunitx}

\usepackage{multirow}
\usepackage{tkz-kiviat}

\usepackage[font=scriptsize]{subcaption}
\usepackage[font=footnotesize]{caption}

\usepackage{mathtools}

\usepackage{dblfloatfix}    
\usepackage{colortbl}

\usepackage{makecell}
\usepackage{diagbox}
\usepackage{tikz-qtree}
\usetikzlibrary{trees} 
\usepackage{cite}
\usepackage{bbold}
\usepackage{bbm}

\usepackage{array}
\newcolumntype{?}{!{\vrule width 1.5pt}}
\newcolumntype{P}[1]{>{\centering\arraybackslash}p{#1}}
\usepackage{makecell}
\usepackage{mdframed}
\usepackage[many]{tcolorbox}
\usepackage{enumitem}

\newtcbox{\mybox}[1][]{nobeforeafter,math upper,tcbox raise base,
  enhanced,frame hidden,boxrule=0pt,interior style={top color=green!10!white,
  bottom color=green!10!white,middle color=green!50!yellow},
  fuzzy halo=1pt with green,drop large lifted shadow,#1}

\usepackage{siunitx}
\usetikzlibrary{fadings}

\tikzfading[name=middle,
            top color=transparent!100,
            bottom color=transparent!100,
            middle color=transparent!20]

\usetikzlibrary{arrows,automata,calc,shapes, positioning,shadows,shadows.blur,shapes.geometric}

\newacronym{3gpp}{3GPP}{3rd Generation Partnership Project}
\newacronym{adc}{ADC}{Analog to Digital Converter}
\newacronym{5g}{5G}{5th generation}
\newacronym{6g}{6G}{6th generation}
\newacronym{aimd}{AIMD}{Additive Increase Multiplicative Decrease}
\newacronym{am}{AM}{Acknowledged Mode}
\newacronym{amc}{AMC}{Adaptive Modulation and Coding}
\newacronym{aqm}{AQM}{Active Queue Management}
\newacronym{awgn}{AGWN}{Additive White Gaussian Noise}
\newacronym{balia}{BALIA}{Balanced Link Adaptation}
\newacronym{bdp}{BDP}{Bandwidth-Delay Product}
\newacronym{bf}{BF}{beamforming}
\newacronym{cc}{CC}{Congestion Control}
\newacronym{cdf}{CDF}{Cumulative Distribution Function}
\newacronym{cn}{CN}{Core Network}
\newacronym{cqi}{CQI}{Channel Quality Information}
\newacronym{cp}{CP}{Control Plane}
\newacronym{csirs}{CSI-RS}{Channel State Information - Reference Signal}
\newacronym{dc}{DC}{Dual Connectivity}
\newacronym{rb}{RB}{Resource Block}
\newacronym{dce}{DCE}{Direct Code Execution}
\newacronym{dci}{DCI}{Downlink Control Information}
\newacronym{udp}{UDP}{User Datagram Protocol}
\newacronym{dl}{DL}{Downlink}
\newacronym{dmr}{DMR}{Deadline Miss Ratio}
\newacronym{dmrs}{DMRS}{DeModulation Reference Signal}
\newacronym{e2e}{E2E}{End-to-End}
\newacronym{ppp}{PPP}{Poission Point Process}
\newacronym{si}{SI}{Study Item}
\newacronym{ecn}{ECN}{Explicit Congestion Notification}
\newacronym{edf}{EDF}{Earliest Deadline First}
\newacronym{enb}{eNB}{eNodeB}
\newacronym{epc}{EPC}{Evolved Packet Core}
\newacronym{es}{ES}{Edge Server}
\newacronym{cav}{CAV}{Connected and Autonomous Vehicle}
\newacronym{fdma}{FDMA}{Frequency Division Multiple Access}
\newacronym{fdd}{FDD}{Frequency Division Duplexing}
\newacronym{upa}{UPA}{Uniform Planar Array}
\newacronym[firstplural=Radio Access Technologies (RATs)]{rat}{RAT}{Radio Access Technology}
\newacronym[firstplural=Radio Access Technology (RTs)]{rt}{RT}{Radio Technology}
\newacronym{fs}{FS}{Fast Switching}
\newacronym{isd}{ISD}{inter-site distance}
\newacronym{ftp}{FTP}{File Transfer Protocol}
\newacronym{gnb}{gNB}{Next Generation Node Base}
\newacronym{harq}{HARQ}{Hybrid Automatic Repeat reQuest}
\newacronym{hetnet}{HetNet}{Heterogeneous Network}
\newacronym{hh}{HH}{Hard Handover}
\newacronym{hol}{HOL}{Head-of-Line}
\newacronym{ia}{IA}{Initial Access}
\newacronym{imt}{IMT}{International Mobile Telecommunication}
\newacronym{iot}{IoT}{Internet of Things}
\newacronym{los}{LOS}{Line of Sight}
\newacronym{lte}{LTE}{Long Term Evolution}
\newacronym{m2m}{M2M}{Machine to Machine}
\newacronym{mac}{MAC}{Medium Access Control}
\newacronym{mc}{MC}{Multi-Connectivity}
\newacronym{mcs}{MCS}{Modulation and Coding Scheme}
\newacronym{mec}{MEC}{Mobile Edge Cloud}
\newacronym{mi}{MI}{Mutual Information}
\newacronym{mimo}{MIMO}{Multiple Input Multiple Output}
\newacronym{mmwave}{mmWave}{millimeter wave}
\newacronym{mptcp}{MPTCP}{Multipath TCP}
\newacronym{mr}{MR}{Maximum Rate}
\newacronym{mss}{MSS}{Maximum Segment Size}
\newacronym{mtd}{MTD}{Machine-Type Device}
\newacronym{mtu}{MTU}{Maximum Transmission Unit}
\newacronym{nfv}{NFV}{Network Function Virtualization}
\newacronym{vnf}{VNF}{ Virtualization Network Function}
\newacronym{sdn}{SDN}{Software Defined Networking}
\newacronym{nlos}{NLOS}{Non Line of Sight}
\newacronym{nlosb}{NLOSb}{Building Non Line of Sight}
\newacronym{nlosv}{NLOSv}{Vehicle Non Line of Sight}
\newacronym{nr}{NR}{New Radio}
\newacronym{ofdm}{OFDM}{Orthogonal Frequency Division Multiplexing}
\newacronym{pdcch}{PDCCH}{Physical Downlonk Control Channel}
\newacronym{pdcp}{PDCP}{Packet Data Convergence Protocol}
\newacronym{pdsch}{PDSCH}{Physical Downlink Shared Channel}
\newacronym{pdu}{PDU}{Packet Data Unit}
\newacronym{pf}{PF}{Proportional Fair}
\newacronym{pgw}{PGW}{Packet Gateway}
\newacronym{phy}{PHY}{Physical}
\newacronym{pbch}{PBCH}{Physical Broadcast Channel}
\newacronym[plural=\gls{mme}s,firstplural=Mobility Management Entities (MMEs)]{mme}{MME}{Mobility Management Entity}
\newacronym{prb}{PRB}{Physical Resource Block}
\newacronym{pss}{PSS}{Primary Synchronization Signal}
\newacronym{pucch}{PUCCH}{Physical Uplink Control Channel}
\newacronym{pusch}{PUSCH}{Physical Uplink Shared Channel}
\newacronym{rach}{RACH}{Random Access Channel}
\newacronym{ran}{RAN}{Radio Access Network}
\newacronym{red}{RED}{Random Early Detection}
\newacronym{rf}{RF}{Radio Frequency}
\newacronym{rlc}{RLC}{Radio Link Control}
\newacronym{rlf}{RLF}{Radio Link Failure}
\newacronym{rrc}{RRC}{Radio Resource Control}
\newacronym{rrm}{RRM}{Radio Resource Management}
\newacronym{rr}{RR}{Round Robin}
\newacronym{rs}{RS}{Remote Server}
\newacronym{rsrp}{RSRP}{Reference Signal Received Power}
\newacronym{rss}{RSS}{Received Signal Strength}
\newacronym{rtt}{RTT}{Round Trip Time}
\newacronym{rw}{RW}{Receive Window}
\newacronym{rx}{RX}{Receiver}
\newacronym{sa}{SA}{standalone}
\newacronym{sack}{SACK}{Selective Acknowledgment}
\newacronym{sap}{SAP}{Service Access Point}
\newacronym{sch}{SCH}{Secondary Cell Handover}
\newacronym{scoot}{SCOOT}{Split Cycle Offset Optimization Technique}
\newacronym{sdma}{SDMA}{Spatial Division Multiple Access}
\newacronym{sinr}{SINR}{Signal to Interference plus Noise Ratio}
\newacronym{sm}{SM}{Saturation Mode}
\newacronym{snr}{SNR}{Signal to Noise Ratio}
\newacronym{son}{SON}{Self-Organizing Network}
\newacronym{ss}{SS}{Synchronization Signal}
\newacronym{srs}{SRS}{Sounding Reference Signal}
\newacronym{sss}{SSS}{Secondary Synchronization Signal}
\newacronym{tb}{TB}{Transport Block}
\newacronym{tcp}{TCP}{Transmission Control Protocol}
\newacronym{tdd}{TDD}{Time Division Duplexing}
\newacronym{tdma}{TDMA}{Time Division Multiple Access}
\newacronym{tfl}{TfL}{Transport for London}
\newacronym{tm}{TM}{Transparent Mode}
\newacronym{prr}{PRR}{Packet Reception Ratio}
\newacronym{trp}{TRP}{Transmitter Receiver Pair}
\newacronym{tti}{TTI}{Transmission Time Interval}
\newacronym{ttt}{TTT}{Time-to-Trigger}
\newacronym{tx}{TX}{Transmitter}
\newacronym{ue}{UE}{User Equipment}
\newacronym{ul}{UL}{Uplink}
\newacronym{uml}{UML}{Unified Modeling Language}
\newacronym{um}{UM}{Unacknowledged Mode}
\newacronym{utc}{UTC}{Urban Traffic Control}
\newacronym{vm}{VM}{Virtual Machine}
\newacronym{rsrq}{RSRQ}{Reference Signal Received Quality}
\newacronym{rssi}{RSSI}{Received Signal Strength Indicator}
\newacronym{crs}{CRS}{Cell Reference Signal}
\newacronym{v2v}{V2V}{Vehicle-to-Vehicle}
\newacronym{v2i}{V2I}{Vehicle-to-Infrastructure}
\newacronym{v2n}{V2N}{Vehicle-to-Network}
\newacronym{v2x}{V2X}{Vehicle-to-Everything}
\newacronym{vn}{VN}{Vehicular Node}
\newacronym{dsrc}{DSRC}{Dedicated Short Range Communication}
\newacronym{ci}{CI}{context information}
\newacronym{voi}{VoI}{value of information}
\newacronym{gps}{GPS}{Global Positioning System}
\newacronym{qos}{QoS}{Quality of Service}
\newacronym{qoe}{QoE}{Quality of Experience}
\newacronym{ml}{ML}{Machine Learning}
\newacronym{ahp}{AHP}{Analytic Hierarchy Process}
\newacronym{lidar}{LIDAR}{Light Detection and Ranging}
\newacronym{sumo}{SUMO}{Simulation of Urban MObility}
\newacronym{wave}{WAVE}{Wireless Access in Vehicular Environment}
\newacronym{c-its}{C-ITS}{Connected Intelligent Transportation System}
\newacronym{dash}{DASH}{Dynamic Adaptive Streaming over HTTP}
\newacronym{http}{HTTP}{HyperText Transfer Protocol}
\newacronym{nt}{NT}{non-terrestrial}
\newacronym{ntc}{NTC}{non-terrestrial communication}
\newacronym{ntn}{NTN}{non-terrestrial network}
\newacronym{haps}{HAPS}{High Altitude Platform Station}
\newacronym{hap}{HAP}{High Altitude Platform}
\newacronym{leo}{LEO}{Low Earth Orbit}
\newacronym{meo}{MEO}{Medium Earth Orbit}
\newacronym{geo}{GEO}{Geostationary Earth Orbit}
\newacronym{uav}{UAV}{Unmanned Aerial Vehicle}
\newacronym{nsat}{nSAT}{Nanosatellite}
\newacronym{ehf}{EHF}{extremely high-frequency}
\newacronym{ioe}{IoE}{Internet of Everyone}
\newacronym{gan}{GaN}{Gallium Nitride}

\makeglossaries

\begin{document}
\pagenumbering{gobble}


\title{Non-Terrestrial Networks in the 6G Era: \\Challenges and Opportunities}

\author{{{Marco Giordani},~\IEEEmembership{Member, IEEE},  {Michele Zorzi},~\IEEEmembership{Fellow, IEEE}}

\thanks{Marco Giordani and Michele Zorzi are with the Department of Information Engineering, University of Padova, Padova, Italy (email: \{giordani, zorzi\}@dei.unipd.it).}}

\maketitle

\begin{abstract}
Many organizations recognize non-terrestrial networks (NTNs) as a key component to provide cost-effective and high-capacity connectivity in future 6th generation (6G) wireless networks.
Despite this premise, there are still many questions to be answered for proper network design, including those associated to latency and coverage constraints. 
In this paper, after reviewing research activities on NTNs, we present the characteristics and enabling technologies of NTNs in the 6G landscape and shed light on the challenges in the field that are still open for future research. 
As a case study, we  evaluate the performance of an NTN scenario in which aerial/space vehicles use millimeter wave (mmWave) frequencies to provide access connectivity to on-the-ground mobile terminals as a function of different networking~configurations.
\end{abstract}

		\begin{tikzpicture}[remember picture,overlay]
\node[anchor=north,yshift=-10pt] at (current page.north) {\parbox{\dimexpr\textwidth-\fboxsep-\fboxrule\relax}{
\centering\footnotesize This paper has been accepted for publication in IEEE Network Magazine, \textcopyright 2020 IEEE.\\
Please cite it as: M. Giordani and M. Zorzi, "Non-Terrestrial Networks in the 6G Era: Challenges and Opportunities," \\ in IEEE Network, vol. 35, no. 2, pp. 244-251, Mar. 2021.}};
\end{tikzpicture}

\section{Introduction} 

While network operators have already started deploying commercial \gls{5g} cellular networks, the research community is  discussing use cases, requirements, and enabling technologies towards \gls{6g} systems~\cite{giordani2020toward}. 
Among other challenges,  current networks fall short of providing adequate broadband coverage to rural regions~\cite{yaacoub2020key}. Moreover, even in the most technologically advanced countries, existing cellular infrastructures may lack the level of reliability, availability, and responsiveness requested by future wireless applications, and show vulnerability to natural disasters. Connectivity outages during natural disasters, in particular, may slow down or impede appropriate reaction, create significant damage to business and property, and even loss of lives.

One solution to increase network resiliency  would be to densify cellular sites, which however involves prohibitive deployment and operational expenditures for network operators and requires high-capacity backhaul connections~\cite{polese2020integrated,yaacoub2020key}.
Moreover, network deployment in rural areas (i.e., the most under-connected areas) is further complicated by the varying degree of terrain that may be encountered when installing  cables or fibers between cellular stations.
Network densification will also inevitably lead to an energy crunch with serious economic and environmental concerns.

To address these issues, 6G research is currently focusing on the development of \glspl{ntn} to promote ubiquitous and high-capacity global connectivity~\cite{mozaffari2018beyond}.
While previous wireless generation networks have been traditionally designed to provide connectivity for a quasi bi-dimensional space, 6G envisions a three-dimensional (3D) heterogeneous architecture in which terrestrial infrastructures are complemented by non-terrestrial stations including \glspl{uav}, \glspl{hap}, and satellites~\cite{giordani2020satellite}.
Not only can these elements provide  on-demand cost-effective coverage  in crowded and unserved areas, but they can also guarantee trunking, backhauling, support for high-speed mobility, and high-throughput hybrid multiplay services.
Notably, the potential of \glspl{ntn} has been acknowledged in the standard activities. A work item for 3GPP Rel-17 has indeed been approved in December 2019 to define and evaluate solutions in the field of \glspl{ntn} for NR, with a priority on satellite access. Study items have also been identified for Rel-18 and Rel-19, thus acknowledging long-term research within the timeframe of 6G.

Research studies on \gls{ntn} are not only limited to 3GPP reports.
For instance, Babich \emph{et al.} presented a novel network architecture for an integrated nanosatellite-5G system operating in the \gls{mmwave} domain~\cite{babich2019nanosatellite}, while in our previous work \cite{giordani2020satellite} we identified the most promising configuration(s) for satellite networking and discussed some design trade-offs in this domain. UAVs were also considered as a tool to complement terrestrial connectivity in critical scenarios~\cite{boschiero2020coverage}. 
Additionally, there currently exist several case studies of \gls{ntn} deployments in different countries, in addition to efforts by international foundations and initiatives~\cite{yaacoub2020key}.

\begin{figure*}[t!]
 \centering
\includegraphics[width=.99\textwidth]{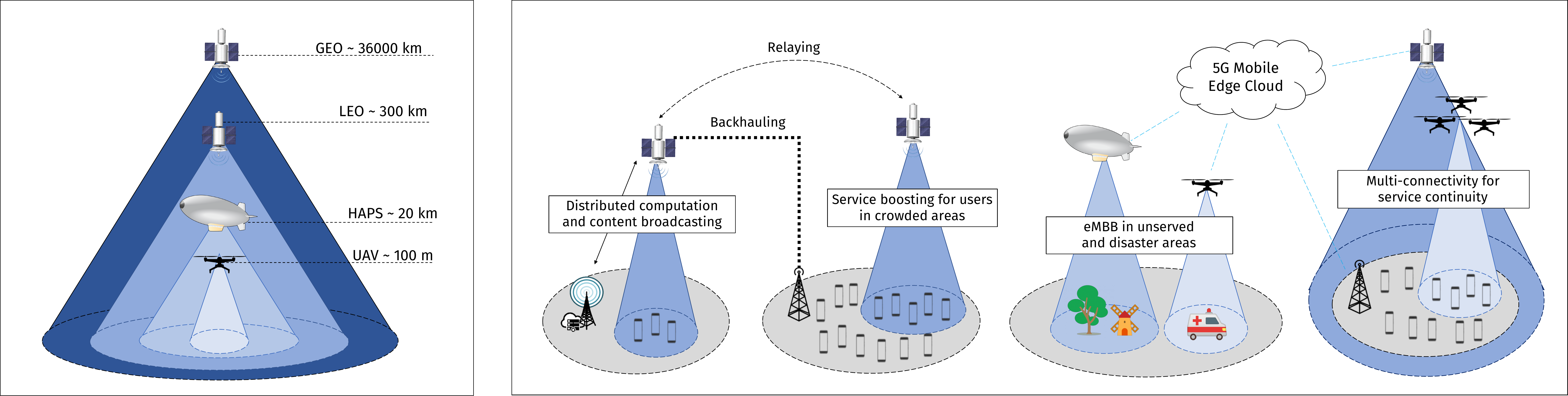}
 \caption{\label{fig:scenario} Non-terrestrial stations (left) and use cases enabled by the integration of terrestrial and non-terrestrial networks (right). }
 \end{figure*}

Nevertheless, despite such earlier investigations, there are still several questions to be answered for proper network design. In particular, while some prior work typically focuses on standalone aerial/space architectures, a formalization of the challenges and opportunities pertaining to a multi-layered network, in which heterogeneous non-terrestrial stations co-operate at different altitudes in an integrated fashion, has not yet been provided. 
Some other  articles, e.g.,~\cite{liu2018space}, review how to improve the protocol stack design in a space-air-ground integrated network, but do not thoroughly explore the most recent technological advancements to achieve performance optimization.
Moreover,  a complete description of \gls{ntn} enabling technological solutions and future research directions is currently scattered in several technical reports, which makes them confusing and tiresome to follow  without the proper~background. 

This paper addresses these challenges by formalizing how \glspl{ntn} can be practically deployed to satisfy emerging 6G application requirements. We focus on (i) new architecture advancements in the  aerial/space industry, (ii) novel spectrum technologies,  e.g., operating in the \gls{mmwave} and optical bands, (iii) antenna design advancements, and (iv) transport layer developments.
Moreover, we shed light on the  research challenges associated to \glspl{ntn}, providing a full-stack perspective with considerations related to spectrum usage,  medium access and higher layers, coverage and mobility management constraints, thus stimulating further research on this topic.
Finally, as a case study, we validate  the feasibility of establishing non-terrestrial communication at \glspl{mmwave} to provide access connectivity to terrestrial nodes.

\section{Non-Terrestrial Networks in 6G} 
\label{sec:non_terrestrial_communication_in}

\glspl{ntn} refer to (segments of) networks operating through  an air/spaceborne vehicle for communication.
While the possibility to integrate satellite technologies to provide access connectivity on the ground was first introduced by 3GPP in Rel-15, more recent activities have been promoted for Rel-16 and Rel-17 to define deployment scenarios and parameters, and identify key potential impacts in NR~\cite{38821}.  Specifications may also continue with enhancements in Rel-18 and Rel-19.


Based on this introduction, in Sec.~\ref{ssub:general_architecture} we describe a typical non-terrestrial architecture, while in Sec.~\ref{ssub:use_cases} we present potential use cases and related deployment scenarios.

\subsection{General Architecture} 
\label{ssub:general_architecture}

Non-terrestrial systems feature (i) a terrestrial terminal, (ii) an aerial/space station, which may  operate similarly to a terrestrial base station, (iii) a service link between the terrestrial terminal and the aerial/space station, and (iv) a gateway that connects the non-terrestrial access network to the core network through a feeder link. Different types of stations can be considered, as depicted in Fig.~\ref{fig:scenario} (left).
\smallskip

\emph{\acrlong{uav} (UAV).}
UAVs fly at  low altitudes (e.g., a few hundred meters) and, thanks to their flexibility,  have recently gained increasing attention to provide broadband wide-scale  wireless connectivity during disasters or temporary events, and relay services for terrestrial mobile nodes.
On the one hand, UAVs can be deployed on-demand, thereby promoting energy efficiency compared to always-on fixed terrestrial infrastructures.
On the other hand,  UAVs incur high propulsion energy consumption to maintain and support their movement, thereby posing severe power management constraints.
\smallskip

\emph{\acrlong{hap} (HAP).}
\glspl{hap} operate in the stratosphere at an altitude of around 20 km. Thanks to their quick deployment and geographical coverage of hundreds of kilometers, these elements are  indeed being considered to support ultra-flexible deployment and cost-effective wireless services, without the prohibitive costs of terrestrial infrastructures.
However,  HAPs may suffer from the need for refueling and challenges related to  stabilization in the air.
\smallskip




\emph{Satellites.}
Satellite stations can be classified according to their  orbit characteristics. \gls{geo} satellites orbit on the Earth’s equatorial plane at an altitude of about 35,800 km and, despite the  significant signal propagation delay and attenuation experienced at such long distance,  can cover very large geographical areas and  are continuously visible from terrestrial terminals.
 \gls{leo} and \gls{meo} satellites, instead, orbit at an altitude between 200 and 2,000 km and 2,000 and 35,000 km, respectively, and guarantee better signal strength and lower propagation delay compared to GEO systems. However, these satellites are non-stationary relative to the Earth’s surface and must operate in a constellation to maintain  service~continuity. The 3GPP is promoting different \gls{ntn} architectures depending on the degree of integration among the different air/spaceborne elements~\cite{38821}. Specifically, the 3GPP envisions:
\begin{itemize}
  \item[(1)] a \emph{transparent} satellite-based \gls{ran} architecture in which the satellite repeats the user's signal from the feeder link to the service link and vice versa;
  \item[(2)] a \emph{regenerative} satellite-based \gls{ran} architecture in which the satellite payload implements regeneration of the signals received from the Earth, while also providing inter-satellite connectivity;
  \item[(3)] a \emph{multi-connectivity} architecture involving two transparent \glspl{ran} (either GEO or LEO or a combination thereof), where integration of terrestrial and non-terrestrial access is also supported.
\end{itemize}

\begin{table*}[t!]
\caption{Enabling technologies for non-terrestrial networks.}
\label{tab:technologies}
\centering
\footnotesize
\renewcommand{\arraystretch}{1.2}
  \begin{tabular}{lll}
\toprule
   & Technology &       Advantage     \\ \midrule
 \multirow{7}{*}{Architecture} & Nano/pico satellites & Small component costs, low  latency, low energy consumption  \\
 & Gallium Nitride (GaN) &  Feasible to install, small form-factor and more efficient components  \\
  & Multi-layered networks & Better spatial and temporal coverage by deploying  satellites in different orbits \\
  & Solid-state lithium batteries &   Safe and  efficient  source of power \\
  & Software Defined Networking (SDN) &  Improved flexibility, automation,  agility through  Virtualization Network Functions (VNFs)\\
  & Flexible payloads &  Dynamic adaptation of beam patterns, frequency, and power allocation  \\ 
  & Hybrid payloads   &  Better trade-off between performance and payload complexity\\ \midrule
  \multirow{4}{*}{Spectrum} & Millimeter waves & Feasibility of ultra-fast connections,  antenna gain, spatial isolation and security  \\
  & UWB modulation & Reduced non-linear signal distortion by encoding the transmitted pulse\\ 
  & Cognitive spectrum & Reduced interference through dynamic spectrum utilization in different frequency bands \\
  & Optical communications &  Feasibility of terabits-per-second connections through extreme bandwidth and directivity\\ \midrule
  \multirow{5}{*}{Antenna} & Reconfigurable phased antennas & Reduced power consumption, size and weight  \\
   & Metasurface antennas & Component miniaturization, high directivity, low sidelobes, fine beamwidth control \\
   & Inflatable/fractal antennas & High-directivity in dynamic scenarios \\ 
   & Coherent antenna arrays & Maintainability, scalability, flexibility, robustness to single points of failure\\
   & Multi-beam architectures & High spectrum efficiency through spatial diversity \\ \midrule
 \multirow{2}{*}{Higher layers} & TCP spoofing & Fast TCP full-buffer capacity through  TCP acknowledgements \\
& TCP multiplexing & High performance by splitting  TCP session into multiple data flows\\
\bottomrule
\end{tabular}
\end{table*}

\subsection{Use Cases} 
\label{ssub:use_cases}

For many years, non-terrestrial devices have been considered to support services like home delivery, meteorology, video surveillance, television broadcasting, remote sensing, and navigation.
However, recent technological developments in the aerial/space industry have opened up the way towards integration between terrestrial and non-terrestrial technologies  to enable more advanced use cases, as illustrated in Fig.~\ref{fig:scenario} (right) and summarized below.
\smallskip

\emph{Communication resilience and service continuity.} Non-terrestrial stations can be deployed to assist existing  base stations in providing high-capacity wireless coverage, e.g., in hot-spot areas or when terrestrial infrastructures are  overloaded. 
Non-terrestrial elements can also provide a secondary backup route to preserve the connection when the primary path is unavailable, e.g., in rural areas or oceans, or when terrestrial towers are  out of service, e.g.,  after natural disasters.
Additionally, these elements can provide on-demand extra capacity  to cell-edge users, the most resource-constrained network entities, thereby promoting fairness in the network.
Finally, aerial platforms can host \gls{mec} functionalities to offer on-the-ground terminals additional computing and storage capabilities, thereby evolving coverage towards 3D. Even though the limited energy support from battery may render the \gls{mec} environment challenging, machine-learning-assisted  migration and technologies for renewable energy production/harvesting and storage are studied to minimize power consumption~\cite{zhou2020mobile}.
\smallskip

\emph{Global satellite overlay.}
When the distance between two terrestrial infrastructures increases, inter-site connectivity through optical fiber may become too expensive. 
A constellation of satellites, where each spacecraft is interconnected with other neighboring spacecrafts via inter-satellite links, can then provide high-capacity access connectivity to on-the-ground devices by relaying the user's signals through an overlay space mesh~network.
\smallskip

\emph{Ubiquitous \gls{iot} broadcasting.}
The wide geographical coverage  and the inherent broadcast nature of aerial/space platforms  make it possible to convey multimedia and entertainment contents to a very large number of user equipments, including in-motion terminals that cannot benefit from terrestrial  coverage like  planes or vessels.
UAVs and satellites can also play the role of moving aggregators for  \gls{iot} traffic, thereby offering global continuity of service for  applications that rely on sensors.
\smallskip

\emph{Advanced backhauling.}
Non-terrestrial terminals can serve on-the-ground backhaul requests wirelessly, e.g.,   for locations where no wired backhaul solutions are available, thereby saving terrestrial resources for the access traffic and  avoiding the costs of traditional fiber-like deployments.
Satellites and other aerial platforms can also complement the terrestrial backhaul in dense regions with high peak traffic demands, thus achieving load balancing.
\smallskip

\emph{Energy-efficient hybrid multiplay.}
Air/spaceborne platforms have the ability to provide high-speed connectivity while promoting energy efficiency. On one side, aerial platforms like \glspl{uav}, while consuming significant energy for hovering, can be deployed on demand implementing smart duty cycle control mechanisms, thereby reducing  management costs of always-on fixed terrestrial infrastructures.
On the other side, space platforms like  satellites can be operated by solar panels which provide  efficient, clean, and renewable energy compared to  traditional energy  sources powering terrestrial devices.

\section{Non-Terrestrial Networks: \\Enabling Technologies} 
\label{ssub:enabling_technologies}
The evolution of \glspl{ntn} will be favored by recent technological advancements in the aerial/space industry, as summarized in Table~\ref{tab:technologies} and described in the following subsections.
We focus on the innovations that do not currently fall within the scope of early \gls{5g} standard activities but could flourish in 6G.
\smallskip

\subsection{Architecture advancements}
Space  manufacturers are improving satellite technologies while further reducing the  operational costs for satellite launch, deployment, and maintenance. 
Nano- and pico-satellites in the LEO orbits, in particular, are emerging as game-changing innovations thanks to their reduced  component costs, and low communication latency and energy consumption. 
Moreover, the adoption of the \gls{gan} technologies on satellites allows the use of smaller form factors and more efficient components compared to their silicon counterparts, thereby saving fuel and area on the payload and improving   operational efficiency~\cite{muraro2010gan}.
Today, the commercialization of \gls{gan} products is restricted to military applications, with most 5G devices utilizing silicon wafer substrates, but their adoption in commercial networks may still be realized for 6G.
Additionally, the availability of multi-layered satellite networks, e.g., LEO and GEO constellations, makes it possible to obtain better spatial/temporal~coverage.
Nevertheless, a real integration between terrestrial and non-terrestrial networks still seems far in the future, and standardization activities are scheduled within the timeframe of 6G.

UAV technology has also improved recently. Solid-state lithium batteries, in particular,  make it possible for \glspl{uav} to  work twice as long compared to today's aerial devices, and are being considered as a safer and more efficient alternative compared to standard lithium-ion batteries.
Furthermore, \gls{uav} swarms, combined with \glspl{hap} and satellites, can operate together to support more robust information broadcasting compared to a standalone deployment by adding redundancy against single points of failure in the path.

Architecture optimization is also favored by the transition to \gls{sdn}~\cite{abdelsalam2019implementation} which,  in combination with network slicing, facilitates the deployment and management of \glspl{vnf}  onto the same physical platform. 
Furthering a trend already started in 5G, 6G will contribute to the design of a disaggregated architecture that can operate in view of the competitive nature of the non-terrestrial environment to guarantee  improved flexibility, automation, and agility in the delivery of services to terrestrial terminals.
Satellite payloads can be realized in software to flexibly adapt beam patterns, frequency, and power allocation, and  react to the dynamics foreseen  in future wireless traffic.
Moreover, hybrid payload implementations, in which the burden of signal processing is split between the on-the-ground gateway and the non-terrestrial station, have been recently studied  to achieve better trade-offs between performance and payload~complexity.
\smallskip

\subsection{Spectrum advancements}
\label{ssec:spectrum}
Non-terrestrial devices have typically been operated in the legacy frequency bands below 6 GHz which, however, may not satisfy the boldest data rate requirements of future beyond-5G services.
Capacity issues can be solved by transitioning to high-frequency communications in the \gls{mmwave} and optical bands, where the huge bandwidths available may offer the opportunity of ultra-fast connections.
However, while the adoption of the \gls{mmwave} spectrum is being successful in the 5G market for both cellular and vehicular networks, it is still unclear whether this technology can be used in the non-terrestrial environment. Solutions are being proposed towards the development of new waveforms and modulation schemes, e.g., impulse-based ultra-wideband (UWB) modulation where information is encoded depending of the characteristics of the transmitted pulse, as a viable approach to reduce the non-linear signal distortion typically experienced at high frequencies~\cite{de2015waveform}.
Moreover, cognitive spectrum  techniques may enable dynamic spectrum utilization in different  bands, while minimizing  interference.

Optical wireless technology can also be used in the feeder link to achieve aggregate capacity in the order of terabits-per-second~\cite{kaushal2017optical}.
Optical transceivers, in fact, leverage higher bandwidth and directivity  compared to radio-frequency systems  and consume much less power and mass.
In this context, atmospheric perturbations and interference from sunlight can be mitigated by wavefront correctors and deformable mirrors, which compensate the signal distortion after propagating through the atmosphere, and advanced modulation schemes.
Error control coding also improves the performance of the optical link by making use of Turbo and convolutional codes.
Nevertheless, despite this  potential, standardization bodies have not yet considered inclusion of optical solutions in the NTN standard, and will be targeting beyond-5G use cases.
\smallskip

\subsection{Antenna advancements}
Aerial/space devices can be equipped with reconfigurable phased antennas offering electronic beam-steering to achieve lower power consumption and reduced size and weight compared to typical mechanical antennas. 
Programmable environments enabled by metasurfaces and intelligent structures are another revolutionary element of the 6G ecosystem to realize antenna component miniaturization, improved directivity, low sidelobes and fine beamwidth control~\cite{samii2015technology}. 
Future trends in the antenna domain further suggest the use of inflatable (i.e., made with flexible-membrane materials) and fractal antennas with unique geometrical designs to obtain high directivity in dynamic scenarios.
Additionally, \glspl{uav} and/or nano-satellites (e.g., in the \gls{leo} orbit) can be deployed in  swarms  to obtain a  distributed coherent antenna array to realize extremely narrowbeam transmissions. Such solution offers maintainability and scalability, as elements can be easily arranged without affecting system operations, and  robustness to single points of~failure.

Advanced antenna solutions allow the implementation of multi-beam architectures that send  information to different spots on the ground through a plurality of beams, thereby maximizing spectrum efficiency through spatial diversity.
The multi-beam approach is further favored by operations in the \gls{mmwave} and optical domains, where the wavelength is so small that it becomes practical to build  large antenna arrays in a small space while maximizing antenna gains through beamforming.
\smallskip

\subsection{Higher-layer advancements}
\glspl{ntn} come with their own set of challenges compared to standalone terrestrial systems, which might make standard transmission protocols, including congestion control over \gls{tcp}, less effective.
Network operators have therefore developed  acceleration techniques that make transport protocols perform better.
TCP spoofing, in particular, is used to send false TCP  acknowledgements to  terrestrial terminals from a spoofing entity (or software) nearby, as if they were sent from the  aerial/space station, thereby making it possible for the TCP control mechanism to quickly reach the maximum supported rate.
TCP multiplexing is another solution that converts a single TCP session into multiple  data flows, each of which can adjust its TCP parameters to  match the characteristics of the non-terrestrial connection.\\

\section{Non-Terrestrial Networks: \\ A Case Study} 
\label{sec:section_name}

As a case study, in this section we assess the feasibility of establishing mmWave communications between terrestrial and satellite terminals, possibly through hybrid integration of multiple aerial/space layers. 
This choice was driven by the fact that the use of satellites operating in the mmWave bands, among all the technologies discussed in Sec.~\ref{ssub:enabling_technologies}, currently represents one of the most promising innovations (as already successfully demonstrated in the cellular and vehicular fields) to offer high-capacity broadcasting capability in NTNs. 

In our simulations, a terrestrial terminal communicates with a satellite placed at different altitudes $h$, and we consider different elevation angles $\alpha 	\in \{\ang{10}, \dots, \ang{90}\}$ and propagation scenarios. 
The channel is modeled as described by the 3GPP in~\cite{38821} and summarized in~\cite[Sec. III]{giordani2020satellite}: specifically, the signal undergoes several stages of attenuation due to atmospheric gases and scintillation.
Terrestrial stations are equipped with directional antennas offering a gain $G_{\rm tx} = 39.7$ dBi~\cite{38821} while, for satellite stations, the gain $G_{\rm rx}$ is varied to consider different antenna architectures. Satellite communication leverages a bandwidth $W$ that depends on the carrier frequency $f_c$: we set $W=20$ MHz for $f_c\leq 6$ GHz,   $W=800$ MHz for $6< f_c\leq 60$ GHz, and $W=2$ GHz for $f_c>60$ GHz.


\begin{figure}[t!]
\centering
  \setlength{\belowcaptionskip}{-0.13cm}
    \setlength\fwidth{0.85\columnwidth}
    \setlength\fheight{0.55\columnwidth}
\includegraphics[width=.99\columnwidth]{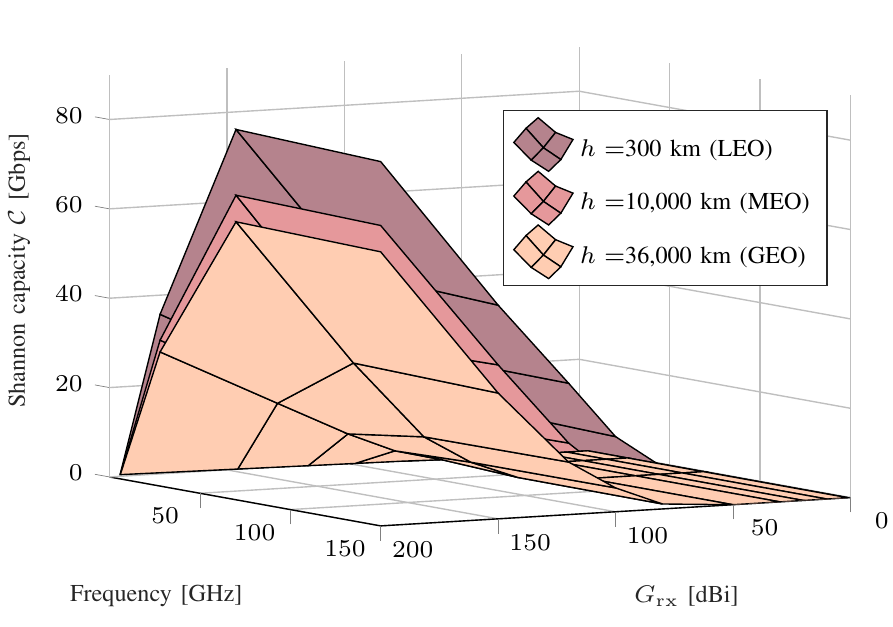}
    \caption{Shannon capacity vs. $h$, $f_c$ and $G_{\rm rx}$, with $\alpha=\ang{10}$ and for a dense urban scenario.}
    \label{fig:surf}
\end{figure}

In Fig.~\ref{fig:surf} we plot the Shannon capacity $\mathcal{C}$, which represents a commonly accepted metric to facilitate accurate benchmarking of wireless networks, as a function of $h$, $f_c$ and $G_{\rm rx}$.
First, we observe that satellite operations in the bandwidth-constrained  below-6 GHz spectrum offer limited capacity (i.e., $< 500$ Mbps), which might be insufficient to satisfy the most demanding beyond-5G use cases.
The performance can be improved by considering \gls{mmwave} transmissions, thanks to the massive bandwidth available at higher frequencies, provided that high-gain directional antennas (i.e., $G_{\rm rx}>50$ dB, as is typical in current satellite antenna technologies) are employed, to cover very long transmission distances.
Fig.~\ref{fig:surf} also makes the case that further increasing $f_c$ beyond 70 GHz would decrease the Shannon capacity due to the increasingly harsh impact of atmospheric absorption in the higher \gls{mmwave} spectrum.

As expected, $\mathcal{C}$ severely reduces for increasing values of $h$, i.e., transitioning from \gls{leo} to \gls{geo} satellites.
Nevertheless, gigabits-per-second capacities can still be reached if the satellite station forms very sharp beams, thereby boosting the performance through massive beamforming.
This is practically feasible since GEO satellites are stationary relative to the Earth's surface and do not require beam re-alignment.

\begin{table*}[t!]
\caption{Open challenges for non-terrestrial networks.}
\label{tab:challenges}
\centering
\footnotesize
\renewcommand{\arraystretch}{1.2}
  \begin{tabular}{ll}
\toprule
 Open challenge &       Explanation     \\ \midrule
 Channel Modeling & Missing adequate characterization of \gls{mmwave} second order statistics, Doppler, fading, multipath \\\midrule
 Spectrum co-existence & Spectrum sharing is required to provide isolation among different non-terrestrial services \\ \midrule
 \multirow{4}{*}{PHY procedures} &  Design of flexible numerology to compensate for large Doppler shift \\
 & Non-linear payload distortions may complicate signal reception \\
 & Large RTTs increase the response time for ACM scheme \\
 & Large RTTs make it infeasible to operate in TDD  \\\midrule 
 HARQ & Large RTTs may exceed the maximum possible number of HARQ processes \\\midrule 
 Synchronization & Large non-terrestrial station's footprint creates a  differential propagation delay among users in the cell \\\midrule 
 Initial access & Channel dynamics may result in obsolete channel estimates \\\midrule 
 Mobility management & Directionality complicates user tracking, handover, and radio link failure recovery \\\midrule 
 \multirow{4}{*}{Constellation management} & Non-terrestrial stations may need to serve a very large number of users  \\
 & Constellation of non-terrestrial stations  is necessary to maintain ubiquitous service continuity \\
 & High cost of satellite launches complicates deployment of dense constellations \\
 & Wireless coordination among air/spaceborne vehicles complicates constellation management \\\midrule 
 \multirow{3}{*}{Higher-layer design} & Channel dynamics  result in obsolete topology information \\
 & Large RTTs result in longer duration of the slow start phase of TCP \\
 & Channel dynamics result in decreased resource utilization due to sudden drops in the link quality \\\midrule 
 \multirow{3}{*}{Architecture technologies} & Unclear where to distribute SDN planes\\
 & Long RTTs prevent long duration of batteries \\
 & Design of central authority making secure network/communication decisions\\ 
\bottomrule
\end{tabular}
\end{table*}

\begin{figure}[t!]
\centering
   \setlength\fwidth{0.9\columnwidth}
   \setlength\fheight{0.5\columnwidth}
   \includegraphics[width=.99\columnwidth]{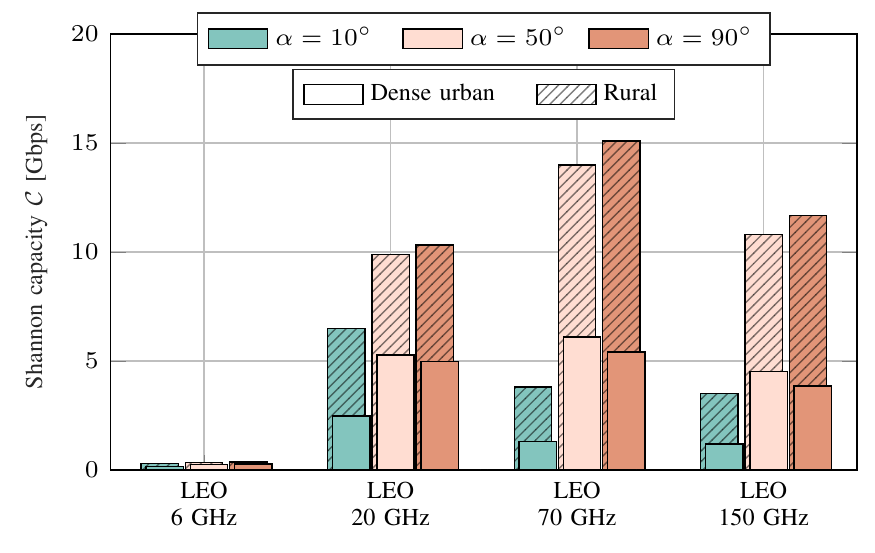}
    \caption{Shannon capacity  vs. $\alpha$ for an  LEO-GND scenario with $h=$300 km, with antenna-gain-to-noise-temperature $G_{\rm rx}/T = 15.9$ dBi/K~\cite{38821}. Dense urban (rural) scenario for  plain (striped) bars. }
    \label{fig:alpha-scenario}
\end{figure}

\begin{figure}[t!]
\centering
  \setlength\fwidth{0.9\columnwidth}
  \setlength\fheight{0.5\columnwidth}
  \includegraphics[width=.99\columnwidth]{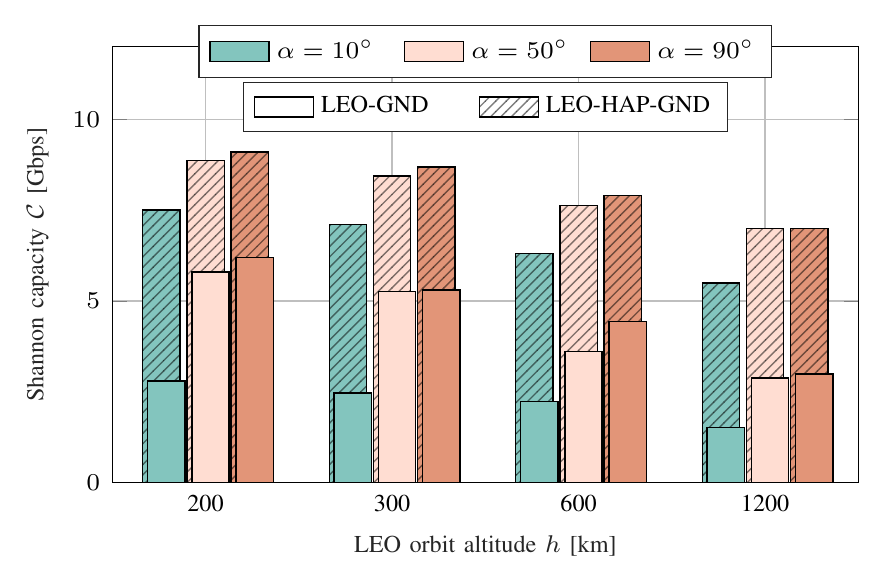}
    \caption{Shannon capacity  vs. $\alpha$ and $h$ for a  dense urban scenario at $f_c=20$ GHz. We compare an LEO-GND (plain bars) and a multi-layered LEO-HAP-GND scenario (striped bars) in which a HAP deployed at 20 km acts as a relay.}
    \label{fig:alpha}
\end{figure}

Similar conclusions can be drawn from Fig.~\ref{fig:alpha-scenario}, where we observe that the system performance decreases at low elevation due to the more severe impact of scintillation absorption (which is caused by sudden changes in the refractive index due to the variation of temperature, water vapor content, and barometric pressure), as the signal has to transit longer through the atmosphere.
Moreover, Fig.~\ref{fig:alpha-scenario} exemplifies that the increased probability of path blockage in the urban scenario may reduce the achievable capacity by more than 60\% at high elevation, compared to a rural scenario.


Despite these promising results, better wireless coverage can be provided when a standalone space layer is assisted by \glspl{hap}  operating in the stratosphere, as  already discussed in Sec.~\ref{ssub:enabling_technologies}. A performance comparison between a standalone LEO scenario (LEO-GND) and a multi-layered scenario (LEO-HAP-GND) in which a HAP bridges the LEO communications towards the ground is plotted in Fig.~\ref{fig:alpha}.
It appears clear that the intermediate HAP offers improved capacity by amplifying the signal from the upstream satellite before forwarding it to the ground, while ensuring a quicker deployment and lower costs compared to spaceborne stations.  The benefits are particularly evident when $h=1,200$ km and $\alpha=\ang{10}$, i.e., when the resulting longer propagation distance from the LEO satellite may deteriorate the signal quality below detectable levels on the ground, with a performance boost of $+250\%$.

\section{Non-Terrestrial Networks: \\ Open Challenges} 
\label{sec:non_ter}

Despite current standardization efforts towards the development of \glspl{ntn}, there remain several open issues for proper protocol design which call for long-term research, as highlighted below and summarized in Table~\ref{tab:challenges}.
\smallskip

\emph{Channel modeling.}
Even though the 3GPP has specified how to characterize \gls{mmwave} propagation for the satellite channel~\cite{38821}, it is currently not investigating second order statistics (including therefore correlation in both space and time), nor the impact of Doppler, fading, and multipath components, which is critical at high frequencies.
Moreover, a general and accurate model of a fully-layered space-air-ground channel is still lacking.
\smallskip

\emph{Spectrum co-existence.}
  As non-terrestrial systems move into the \gls{mmwave} bands, where other systems have been operating for many years (e.g., satellites offering weather forecasting services), consideration needs to be given to the co-existence  among different networks.
  The main challenge is the development of flexible spectrum sharing techniques that maintain adequate isolation among different communications while ensuring reasonable licensing costs.
\smallskip

\emph{PHY procedures.}
In the non-terrestrial case even the highest available sub-carrier spacing in the frame structure may not be enough to compensate for the large Doppler experienced considering the high speed of aerial/space stations.
Moreover, the large propagation delays  in NTNs may create a larger response time for the \gls{amc} scheme loop and requires a margin to compensate for the possible outdated control signals exchanged during channel estimation.
Notably, in an integrated terrestrial/non-terrestrial framework, different network elements on the end-to-end communication path may process the information at different rates, thus contributing to the overall communication delay.
Additionally,   \gls{tdd}, which is frequently considered in terrestrial networks, may be infeasible in non-terrestrial networks since  guard times must be  proportional to the propagation delay.



  
\smallskip

\emph{HARQ.}
  The long \gls{rtt} experienced in non-terrestrial networks may  exceed the maximum possible number of \gls{harq} processes that are typically supported in 5G NR systems. In this regard, simply increasing the number of  processes  may not be feasible due to memory restrictions at the mobile terminal's side.
  Long \glspl{rtt} also require large transmission buffers, and potentially limit the number of retransmissions allowed for each transmission.
\smallskip

\emph{Synchronization.}
	Non-terrestrial systems are fast-moving, and typically feature larger cells compared to terrestrial networks.
	At low  elevation angles, this may create a very large differential propagation delay between users at the cell edge and those at the center (up to 10 ms for GEO satellites~\cite{38821}), thereby raising synchronization issues.
\smallskip

	 \emph{Initial access and channel estimation.}
	Initial access makes on-the-ground terminals establish a physical connection with a non-terrestrial station by detecting synchronization signals.
	This is particularly challenging in non-terrestrial applications, where the channel may vary quickly over time, as the initial estimate may rapidly become obsolete.
  Also, in space-ground integrated networks, each intermediate node tends to associate to a gateway based on its own unilateral benefit, neglecting the potential disadvantages on the whole network performance.
\smallskip

	\emph{Mobility management.}
 	When operating at \glspl{mmwave} to maintain high-capacity connections, directionality is  required in order to achieve sufficient  link budget. In this case, fine alignment of the  beams has severe implications for the design of  control operations, e.g., user tracking, handover,  and radio link failure recovery. 
  These challenges are particularly critical in the non-terrestrial domain, where the very high speed of aerial/space platforms  could result in  loss of beam alignment before a data exchange is completed. The increased Doppler  encountered at high speed  could also make the channel non reciprocal, thus impairing the feedback over a broadcast channel.
\smallskip

	 \emph{Constellation management.}
	 A  non-terrestrial station has a larger footprint than a terrestrial  cell and is required to serve a larger number of on-the-ground terminals. This may result in saturation of the available bandwidth, with strong implications for latency and throughput~performance.

   Additionally, air/spaceborne vehicles  move rapidly relative to the Earth’s surface and may create regions where coverage is not continuously provided. A constellation is thus necessary to maintain ubiquitous service continuity. 
   When configuring multiple satellites that move in different orbits to operate in an integrated fashion, however, constellation management is  hindered by handovers and load balancing among the different~layers.

	 Moreover, while in the terrestrial scenario coordination between base stations is possible through fiber connections  or via a central entity, coordination among  non-terrestrial air/spaceborne stations has to be implemented wirelessly, thus further complicating constellation management.
\smallskip

\emph{Higher-layer design.}
Current network/transport protocols may show low performance when \glspl{ntn} are involved.
First, topology information  may quickly become obsolete (especially considering unpredictable mobility, e.g., for \gls{uav} swarms) and must constantly be refreshed, thus increasing the communication overhead.
Second, a large \gls{rtt} results in a longer duration of the slow start phase of \gls{tcp}, during which the sender may  take inordinately long before operating at full bandwidth.
Third, sudden drops in the link quality, which may be common in \glspl{ntn}, make the sender reduce  its transmission rate, thus leading to a drastic  decrease in resource utilization.
Finally, when a multi-layered integrated network is considered, different network devices may support different (and sometimes conflicting) communication protocols, thus  complicating network management.
\smallskip

\emph{Architecture technologies.}
It is still unclear where to  distribute SDN planes for proper service delivery; the choice depends on different factors, like the available processing  capabilities or the achievable transmission rate.



Furthermore, due to the large distances involved  in non-terrestrial operations and the resulting severe path loss experienced, the transmit power is typically to be set as close as possible to the saturation point. 
This could reduce the duration of batteries, which is particularly critical in scenarios where aerial devices are used to support \gls{iot} applications.

Finally, an integrated space-air-ground architecture should envision the existence of a trusted central authority making secure network topology and  communication decisions to prevent malicious nodes from being selected as a gateway.

\section{Conclusions} 
\label{sec:conclusions}

Non-terrestrial networks are being investigated as a key component of the 6G framework to support global, ubiquitous and continuous connectivity, and to overcome the coverage  limitations of envisioned 5G networks.
In this paper we  overviewed recent  advancements that will make non-terrestrial networks a reality, including the development of new aerial/space architectures, and innovative spectrum and antenna technologies.
As a case study, we demonstrated that the \gls{mmwave} frequencies can be used to establish high-capacity connections between on-the-ground terminals and satellite/HAP gateways, provided that sharp beams are formed.
Despite such promises, we also summarized current open challenges for the deployment of  non-terrestrial networks, thereby stimulating further research in this domain.
Most importantly, our future studies will be dedicated to exploring the relationship between capacity performance and energy efficiency in the non-terrestrial ecosystem.

\section*{Acknowledgments}

Part of this work was supported by the US Army Research Office under Grant no. W911NF1910232 ``Towards Intelligent Tactical Ad hoc Networks (TITAN)''.

\bibliographystyle{IEEEtran}

\vspace{-10.5cm}

\begin{IEEEbiographynophoto}{Marco Giordani}
[M'20] received his Ph.D. in Information Engineering in 2020 from the University of Padova, Italy,  where he is now a postdoctoral researcher and adjunct professor.
He visited  NYU and TOYOTA Infotechnology Center, Inc., USA.
In 2018 he received the “Daniel E. Noble Fellowship Award” from the IEEE Vehicular Technology Society. His research  focuses on protocol design for 5G/6G mmWave cellular and vehicular networks.
\end{IEEEbiographynophoto}%
\vspace{-10.5cm}

\begin{IEEEbiographynophoto}{Michele Zorzi}
[F'07] is with the Information Engineering Department of the University of Padova, focusing on wireless communications research. He was Editor-in-Chief of IEEE Wireless Communications from 2003 to 2005, IEEE Transactions on Communications from 2008 to 2011, and IEEE Transactions on Cognitive Communications and Networking from 2014 to 2018. He served ComSoc as a Member-at-Large of the Board of Governors from 2009 to 2011, as Director of Education and Training from 2014 to 2015, and as Director of Journals from 2020 to 2021.
\end{IEEEbiographynophoto}

\end{document}